\documentclass[twocolumn]{aastex63}

\usepackage{float}
\usepackage{gensymb}
\usepackage{lineno}
\shorttitle{Correlation between A(Li) and surface rotation}
\shortauthors{$\rm Singh\ et\ al.\ 2025$}

\begin{document}

% \title{Internal rotation and A(Li) of low mass evolved stars}
\title{Li-enrichment in red clump giants: Clues for past binary interaction or merger events}

% \correspondingauthor{Raghubar Singh}
% \email{raghubar2015@gmail.com}

\correspondingauthor{Raghubar Singh, Gang Zhao}
\email{raghubar2015@gmail.com, gzhao@nao.cas.cn}
\author[0000-0001-8360-9281]{Raghubar Singh}
\affiliation{National Astronomical Observatories, Chinese Academy of Sciences, 20A Datun Road, Chaoyang District, Beijing, China (100101)}

\author[0000-0001-9246-9743]{Bacham E. Reddy}
\affiliation{Department of Physics, Indian Institute of Technology Jammu, Jammu, 181221, India}

\author[0000-0002-4282-605X]{Anohita Mallick}
\affiliation{Indian Institute of Astrophysics, 560034, 100ft road Koramangala, Bangalore, India}

% \author[0000-0002-4331-1867]{Jeewan C. Pandey}
% \affiliation{Aryabhatta Research Institute of Observational Sciences (ARIES), Manora Peak, Nainital 263001, India}

\author[0000-0002-8980-945X]{Gang Zhao}
\affiliation{National Astronomical Observatories, Chinese Academy of Sciences, 20A Datun Road, Chaoyang District, Beijing, China (100101)}

\begin{abstract}
To understand the underlying mechanisms for high lithium abundances among core He-burning or red clump (RC) giants, we analyzed a sample of 5227 RC giants of mass M $\leq$ 2~M$_{\odot}$ using spectra and asteroseismic data. We found 120 RC giants ($\sim$2~$\%$) with a lower limit of A(Li) = 0.7~dex, a factor of 40 more than their predecessors close to the RGB tip. Of the 120 RC giants,  we could measure actual rotations for 16 RC giants using stellar spots from the Kepler light curve analysis. We found that most of the high rotation RC giants are also very high Li-rich RC giants, and the rotation seems to decline rapidly with Li abundance depletion, suggesting that both the high rotation and high Li abundance are transient phenomena and associated with a single source. Further, we found a significantly high occurrence of 15~$\%$ and 12~$\%$ of Li-rich RC giants among extremely low-mass RC giants and RC giants with anomalous [C/N] ratios, respectively. The extremely low mass, fast rotation and anomalous [C/N] values of RC giants are attributed to their past binary interaction/merger history. The results pose a question of whether the binary interaction/merger is a prerequisite along with the He-flash for Li-enhancement among RC giants. 
\end{abstract}


\section{Introduction}
Understanding the production and evolution of lithium abundance, A(Li) \footnote{A(Li) = 12 + log(N(Li)/N(H)), N represents the number density of respective elements.} in stars is important to gain insights into the stellar interior processes and probably to account for Li-enrichment of the Galaxy. Stellar theoretical models predict the depletion of A(Li) during their evolution along the red giant branch \citep{Iben1967a}. The very high A(Li) seen in 1-2$\%$  of red giants \citep{Brown1989, Deepak2019, Casey2019, Yan2021} has been an unsolved problem since its discovery in a typical red giant \citep{wallerstein1982}. 

Rapid progress has been made in the last decade owing to the availability of large data sets of spectra from surveys like GALAH \citep{DeSilva2015}, LAMOST \citep{Cui2012, Zhao2012} and APOGEE \citep{abdurrouf2022} and the time-resolved photometry from space missions e.g., Kepler \citep{Borucki2010} and TESS \citep{Ricker2015} for asteroseismic analysis \citep{singh2019, Casey2019, Yan2021}. These studies suggest that the Li-rich giants are common among the He-core burning phase (red clump) with an average A(Li) = 0.7~dex which is a factor of about 40 more than their counterparts near the tip of RGB, before the He-flash, for which model predictions and observations agree with an upper limit of A(Li) $\sim$-0.9~dex \citep{Kumar2020}. These results, combined with the results that high A(Li) is only prevalent among RC giants \citep{Mallick2023}, reinforced the hypothesis that the He-flash may be the most likely site for Li-production. Further, the study by \cite{singh2021} has shown that A(Li) decreases steeply with an asymptotic period spacing value, a proxy for time evolution for the transition from a degenerate core to a fully convective He-burning core. The relation suggests that RC giants with high A(Li) have undergone the Li-enrichment process very recently, and they are young compared to RC giants with normal A(Li) ($\leq$ 1.0~dex). 

In general, it is known that lithium gets synthesized via the Cameron-Fowler \citep{Cameron1971} process in which the initial reaction $^{3}$He ($\alpha$, $\gamma$) $^{7}$Be occurs in deeper layers of the He-burning shell and the second reaction $^{7}$Be (e, $\nu$) $^{7}$Li to occur in sufficiently cool regions so that Li has enough time to mix-up with the outer atmosphere. However, the theory of nucleosynthesis and mixing processes during the He-flash is not well understood. A few attempts have been made recently to explain the occurrence of high A(Li) among RCs due to single stars' evolution \citep{Schwab2020} and also due to binary interactions \citep{Zhang2013, Zhang2020, Rui2024}.

The new observational results are emerging, showing a significant number of RC giants with extremely low mass \citep{Li2022}, which are also found to have anomalous carbon-to-nitrogen ([C/N]) ratios \citep{Bufanda2023}. 
% The masses ($\leq$ 0.7~M$_{\odot}$) of the giants are such that they would be much older than the age of the universe even if we account for the expected mass loss (~0.2~M$_{\odot}$) during the evolution of RGB \citep{Miglio2021}.
Models of binary star evolution predict significant mass loss due to the evaporation of a common envelope \citep{Matteuzzi2024}, which influences the [C/N] values \citep{Izzard2018}. The binary interactions may result in high surface rotation \citep{tayar2015} and extremely low mass. In light of these new insights into the properties of red clump giants, we have performed spectroscopic and light curve analysis of a large sample of giants and discussed whether the Li-rich giants correlate with the giants of high rotation, extremely low mass and anomalous [C/N]. 

\section{Sample selection and Analysis}
We collected a sample of 7110 low mass (M $\le$ 2~M$_{\odot}$) red clump giants from published catalogues \citep{Mosser2014, Vrard2016, Yu2018, Gaulme2020}. The mass and evolutionary phase information are based on asteroseismic analysis of Kepler data. We cross-matched the Kepler sample with the Large Sky Area Multi-Object Fiber Spectroscopic Telescope (LAMOST) \citep{Zhao2006, Zhao2012, Cui2012} DR10 publicly available catalogues and found 5227 giants having spectra of low (R = 1800) or medium resolution (R = 7500). The spectra were inspected for detectable Li features at $\lambda$6707~\AA\ and found 120  RC giants with weak to very strong Li lines. The radial velocity, carbon and nitrogen abundances were adopted from the APOGEE DR17 catalogue \citep{abdurrouf2022}. Of the 5227 RC giants in our sample, we found C and N abundances for 3640 RC giants. We also made use of the Gaia DR3 astrometry \citep{Gaia2016, Gaia2018} and radial velocity data to understand the binary status of the sample RC giants. We have subjected the 120 RC giants to the measurement of Li abundances and rotation period. The sample stars have a mass range of  0.52 -- 1.7~M$_{\odot}$ with a mean mass of 1.1~M$_{\odot}$. They are mostly solar metallicity RC giants with a mean value of [Fe/H] = 0.0~dex.
%Abundances were derived using spectrum synthesis analysis. We derived stellar surface rotation from the light curve analysis of stellar spots. 

\noindent
{\bf Li abundance:} Measurement of Li abundance from spectra requires stellar atmospheric parameters (T$_{\rm eff}$, log~g and [M/H]), atomic and molecular data; and radiative transfer code. We used log~g values derived from asteroseismically derived mass and radius. The metallicity and temperature values were adopted from the LAMOST catalogue. We generated 1~D local thermodynamic equilibrium models suiting the sample stars' parameters by interpolating the Kurucz model grids\footnote{\url{https://github.com/kolecki4/PyKMOD}} \citep{CastelliKuruxz2004}. The parameter microturbulance velocity was derived using an empirical relation $ \rm \xi_{t} = (2.13 \pm 0.05) - (0.23 \pm 0.03)$log~g km s$^{-1}$ \citep{Kirby2009}. The atomic and molecular data required for spectrum synthesis were adopted from  {\it Linemake \footnote{\url{https://github.com/vmplacco/linemake}}} code \citep{Placco2021}. The spectra were continuum-normalized and corrected for radial velocity using template spectra in {\it IRAF} \footnote{\url{https://iraf-community.github.io}} \citep{Tody1986}. Model spectra for stars' parameters were generated using the 1D LTE radiative transfer code {\it MOOG \footnote{\url{https://www.as.utexas.edu/~chris/moog.html}}} \citep{Sneden1973} using an appropriate parameter of FWHM for line widths. We obtained Li abundances by matching the model spectra with those of the observed spectra. The Li abundances range from 0.7 - 4.2~dex. The lower limit is due to spectral resolution and the quality of spectra. The measured abundances are corrected for non-LTE effects \citep{LindK2009}. Among the 120 RC giants, we found 30 of them as super Li-rich giants ({A(Li)$\ge$3.3 dex}). Most of them, $\sim$94$\%$, have higher lithium than A(Li) $\sim$ 1.5~dex, the classical lower limit for lithium-rich giants on the RGB. 

\begin{figure}
    \centering
    \includegraphics[width=\columnwidth]{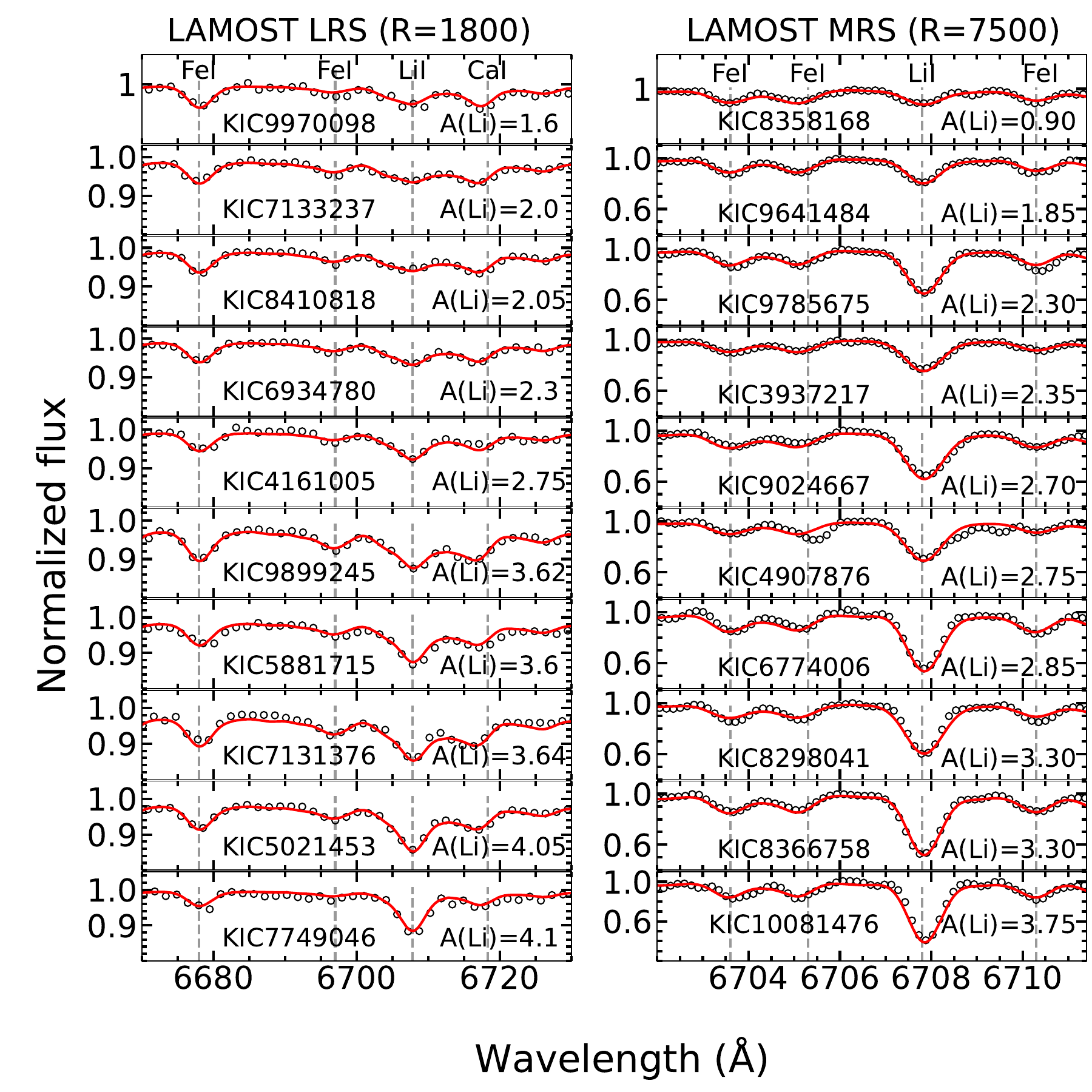}
    \caption{Comparison of computed spectra (red solid line) with the observed spectra (empty circle) around the Li doublet line at 6707.7~\AA\ for a few representative stars for LAMOST spectra of R = 1800 and 7500.}
    \label{fig:lamostsynth}
\end{figure}

\begin{figure}
    \centering
    \includegraphics[width=\columnwidth]{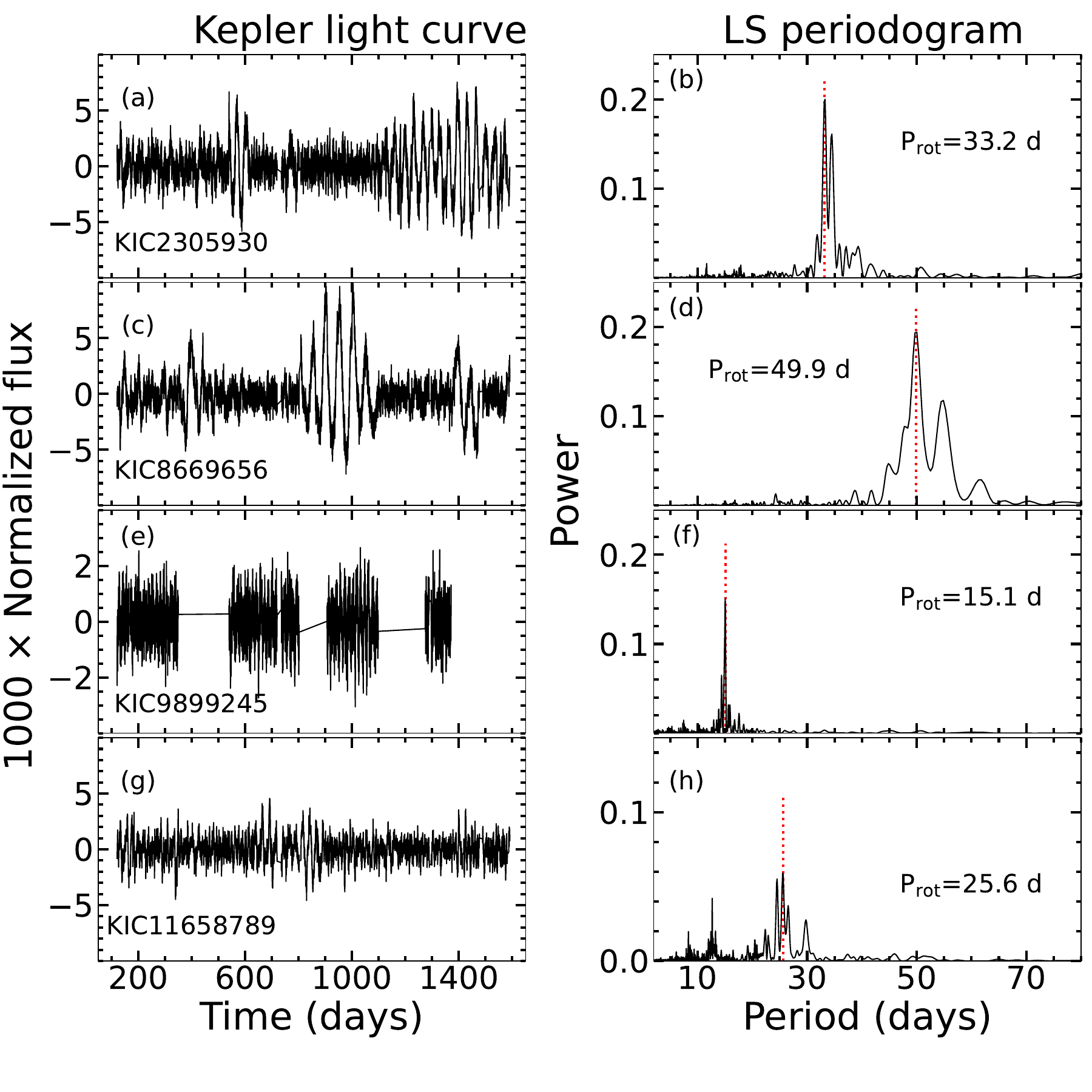}
    \caption{Rotational period estimation from Kepler data. In the left panel long cadence Kepler light curves of four stars. In the right panel corresponding Lomb-Scargle (LS) periodogram of stars. The estimated rotation period, which is the tallest peak in the periodogram, is marked with the red dotted vertical line.}
    \label{fig:keplerlc}
\end{figure}

\noindent
{\bf Rotational velocity measurement of stars:}
Stellar rotation manifests in the form of broadening of spectral lines and quasi-periodic brightness variation in the light curve. Deriving stars' $\rm V_{\rm rot}$ values using spectral line broadening is not possible due to the low spectral resolution (40 -- 160~km s$^{-1}$). Further, the derived values are a function of an inclination angle between the line of sight and the axis of the stars' rotation, for which we don't have information. We made use of the fact that the magnetic activity manifests spots on the stars' surfaces similar to sunspots. The latter has an added advantage for measuring actual rotation if the radii of the stars are obtained.  

We analyzed Kepler long cadence (time = 29.42~m) photometric data of many quarters; a few of the sample giants have 17 quarters, to detect rotation modulation in the flux caused by star spots. The Q0, Q1 and Q17 quarters have shorter lengths, so we excluded them from the analysis. The presence of spots provides sinusoidal fluctuations in the stellar brightness. Lightcurve data for stars is retrieved from {\it MAST \footnote{\url{ https://mast.stsci.edu/portal/Mashup/Clients/Mast/Portal.html}}} using lightkurve package \footnote{\url{https://github.com/lightkurve/lightkurve?tab=readme-ov-file}} \citep{lightkurve2018}. The data from individual quarters were stitched together for each star's light curve after normalization. We followed Lomb-Scargle (LS) periodogram \citep{Lomb1976, Scargle1982} method to estimate the rotational period of the stars. The periodogram of the light curve with spot modulations shows a large amplitude peak at a short frequency. We only chose stars with clear peaks with high power ($>$0.01) corresponding to brightness variation in Fourier spectra.  We selected the highest peak as a rotational period of the star, see Fig.~\ref{fig:keplerlc}. We searched for rotation periods ranging from 1 to 150 days. Uncertainty in the rotational period is measured by fitting Gaussian to the high amplitude low-frequency peak, see Figure~\ref{fig:keplerlc}. The equatorial rotation velocity (V$_{\rm rot}$) of stars is measured  by $\rm V_{rot} = (2 \pi R)/P_{rot}$. Uncertainty in the derived rotation values stems from the uncertainty in the measurement of radius and rotation period. The extracted period ranges from 16 to 60 days, see Table~\ref{tab:tabl1}. With the derived precise asteroseismic radii of stars, we could derive the $\rm V_{\rm rot}$ values in the range of 10 -- 40~km s$^{-1}$. 

\begin{table*}[]
    \centering 
    \begin{tabular}{ccccccccc}
    \hline
    Star & T$_{\rm eff}$ & log~g & $\rm [Fe/H]$& A(Li)&M & R & P$_{\rm rot}$ & V$_{\rm rot}$\\
    KIC & K & cm s$^{-2}$ & dex& dex&M$_{\odot}$ & R$_{\odot}$ & days & km s$^{-1}$\\
    \hline
2305930 &4858$\pm$42 &2.35$\pm$0.02 &-0.47$\pm$0.03 &3.91$\pm$0.2 &0.72$\pm$0.12 &9.44$\pm$0.72 &33$\pm$1.1 & 14.5.5$\pm$1.2\\ 
3534438 &4999$\pm$38 &2.43$\pm$0.02 &-0.06$\pm$0.03 &3.95$\pm$0.15 &0.52$\pm$0.10 &7.28$\pm$0.1 &15$\pm$.5 & 24.6$\pm$0.88\\
3937217 &4810$\pm$53 &2.38$\pm$0.02 &-0.21$\pm$0.04 &3.0$\pm$0.1 &0.97$\pm$0.09 &10.78$\pm$0.39 &56.9$\pm$3.1 & 9.6$\pm$0.62\\ 
4743066 &4811$\pm$27 &2.36$\pm$0.02 &-0.18$\pm$0.02 &2.35$\pm$0.3 &1.06$\pm$0.26 &11.62$\pm$72 &50.51$\pm$3.1 & 11.6$\pm$1.01\\ 
5021453 &4754$\pm$34 &2.41$\pm$0.01 &-0.07$\pm$0.02 &4.05$\pm$0.12 &1.02$\pm$0.14 &10.48$\pm$0.58 &30.35$\pm$0.7 & 17.5$\pm$1.04\\ 
% 5943345m &4934$\pm$92 &2.64$\pm$0.08 &0.01$\pm$0.05 &3.9$\pm$0.2 &2.62$\pm$0.31 &10.48$\pm$0.58 &33$\pm$1.1 & 20.5$\pm$1.5\\ 
% 7547646 &5027$\pm$36 &2.81$\pm$0.05 &0.13$\pm$0.03 &3.9$\pm$0.2 &2.74$\pm$0.21 &12.96$\pm$0.55 &33$\pm$1.1 & 20.5$\pm$1.5\\ 
5879876 & 4618$\pm$100 & 2.37$\pm$0.03&-0.6$\pm$0.3& 3.8$\pm$0.2 &0.93$\pm$0.23&10.4$\pm$0.97 & 45$\pm$5 & 11.7$\pm$1.7\\ 
7899597 &4715$\pm$37 &2.41$\pm$0.02 &-0.09$\pm$0.03 &3.55$\pm$0.08 &1.28$\pm$0.23 &11.76$\pm$0.95 &50$\pm$3.1 & 11.90$\pm$1.2\\ 
% 8358168m &4958$\pm$39 &2.69$\pm$0.05 &-0.01$\pm$0.03 &2.1$\pm$0.3 &0.68$\pm$0.09 &8.64$\pm$0.41 &20$\pm$0.7& 21.3$\pm$1.3\\ 
% 8378545m &4974$\pm$43 &2.74$\pm$0.05 &0.18$\pm$0.03 &3.9$\pm$0.2 &4.3$\pm$1.09 &17.06$\pm$1.47 &33$\pm$1.1 & 20.5$\pm$1.5\\ 
8869656 &4775$\pm$30 &2.40$\pm$0.01 &-0.31$\pm$0.03 &3.61$\pm$0.07 &1.09$\pm$0.09 &10.91$\pm$0.37 &50$\pm$2 & 11.04$\pm$0.6\\ 
8879518 &4778$\pm$47 &2.57$\pm$0.01 &0.01$\pm$0.03 &3.51$\pm$0.08 &1.61$\pm$0.11 &10.89$\pm$0.29 &40.1$\pm$0.5 & 13.8$\pm$0.5\\ 
9364778 &5000$\pm$22 &2.45$\pm$0.01 &-0.16$\pm$0.01 &3.43$\pm$0.03 &1.7$\pm$0.14 &12.93$\pm$0.44 &35.6$\pm$0.3 & 18.4$\pm$0.7\\ 
% 9657730 & 5196$\pm$161 &2.40$\pm$0.03& -0.2$\pm$0.3& 3.4$\pm$0.1 &0.62$\pm$0.08&8.23$\pm$0.41& 18$\pm$2 & 2p3$\pm$3\\         
9833651 &4678$\pm$32 &2.48$\pm$0.01 &0.14$\pm$0.03 &3.41$\pm$0.11 &1.27$\pm$0.10 &10.75$\pm$0.33 &47.4$\pm$3.06 & 11.5$\pm$0.9\\ 
9843104 &4665$\pm$38 &2.36$\pm$0.02 &-0.02$\pm$0.03 &2.5$\pm$0.25 &1.16$\pm$0.21 &11.78$\pm$0.73 &27$\pm$3.1 & 22.1$\pm$2.8\\ 
9899245 &4700$\pm$29 &2.44$\pm$0.03 &0.09$\pm$0.02 &3.7$\pm$0.4 &1.435$\pm$0.28 &11.99$\pm$0.8 &15$\pm$0.3 & 40.5$\pm$2.8\\ 
% 9970098 &4990$\pm$32 &2.35$\pm$0.04 &-0.76$\pm$0.02 &2.0$\pm$0.25 &1.13$\pm$0.10 &10.61$\pm$0.35 &27$\pm$2.3 & 19.9$\pm$1.8\\ 
10081476 &4457$\pm$42 &2.33$\pm$0.01 &0.16$\pm$0.03 &3.8$\pm$0.3 &1.17$\pm$0.18 &12.3$\pm$0.85 &20$\pm$2 & 31.1$\pm$3.8\\ 
11658789 &4995$\pm$41 &2.37$\pm$0.03 &-0.69$\pm$0.04 &3.9$\pm$0.1 &0.87$\pm$0.14 &9.51$\pm$0.59 &25.55$\pm$0.2 & 18.8$\pm$1.2\\ 
% 12689984 &4990$\pm$27 &2.68$\pm$0.04 &0.07$\pm$0.02 &3.9$\pm$0.2 &2.67$\pm$0.27 &13.23$\pm$0.49 &47.52$\pm$0.5 & 20.5$\pm$1.5\\ 
11129153 &4830$\pm$100 &2.42$\pm$0.02 &-0.37$\pm$0.15 &1.45$\pm$0.2 &0.96$\pm$0.09 &10.84$\pm$0.39 &44$\pm$4.1 & 12.5$\pm$1.25\\
% 3751167 &4914$\pm$80 &2.33$\pm$0.27 &-0.76$\pm$0.15 &3.4$\pm$0.12 &0.95$\pm$0.22 &11.04$\pm$1.02 &10.9$\pm$2.1 & 51$\pm$11\\
% 9094309 &5092$\pm$161 &2.44$\pm$0.01 &-0.37$\pm$0.30 &4.0$\pm$0.17 &1.23$\pm$0.12 &11.04$\pm$0.45 &11$\pm$3.1 & 51$\pm$14\\
\hline 
    \end{tabular}
    \caption{Stellar parameters of the rapidly rotating Li-rich giants stars.}
    \label{tab:tabl1}
\end{table*}

\section{Results and discussion}
The derived rotational velocities, along with A(Li), and values of mass for the 16 RC giants are given in Table~\ref{tab:tabl1}. The values of V$_{\rm rot}$ are well above the typical red giant rotation velocity \citep{deMadeiros1996}. The values of $\rm V_{\rm rot}$ for RC giants are expected to be low even after taking into account the expected spin-up of stars due to the sudden drop in the size of the red giants after the He-flash and mass loss \citep{Sills2000, Casey2019, Singh2024}. For giants with M $<$ 1.1~M$_{\odot}$ the expected rotation is quite low and probably undetectable whereas stars with mass 1.1 $<$ M(M$_{\odot}$) $<$ 1.7 are expected to have a maximum rotation of $\rm  V_{\rm rot}$ = 10~km s$^{-1}$ at red clump phase \citep{tayar2015}. We have RC giants with a wide range of masses, and nine of them are below $\sim$1.1~M$_{\odot}$, and seven of them have a mass between 1.1 $<$ M (M$_{\odot}$) $<$1.7. All the 16 RC giants have $\rm V_{\rm rot}$ $>$ 10~km s$^{-1}$, falling in the category of rapid rotators. 

This is the largest sample of Li-rich RC giants for which $V_{\rm rot}$ values were measured based on stellar spots. The higher surface rotation may be due to internal processes or external events. The internal process is related to He-flash and the consequent stars' transition from the RGB tip to the RC phase. Post He-flash, stars undergo a sudden decrease in size of the stars which may increase the stars' surface rotation. This aspect has been probed by \cite{Singh2024} study which found that the estimated surface rotation values for red clump giants do not exceed $V_{\rm rot}$ $=$ 7~km s$^{-1}$.
% They found the estimated values not exceeding $V_{\rm rot}$ $=$ 7~km s$^{-1}$ at RC. 
Further, the suggestion that the highly rotating central core can translate its angular momentum into higher surface rotation is not found, as these values are too small. 

\begin{figure}[!ht]
    \centering
    \includegraphics[width=\columnwidth]{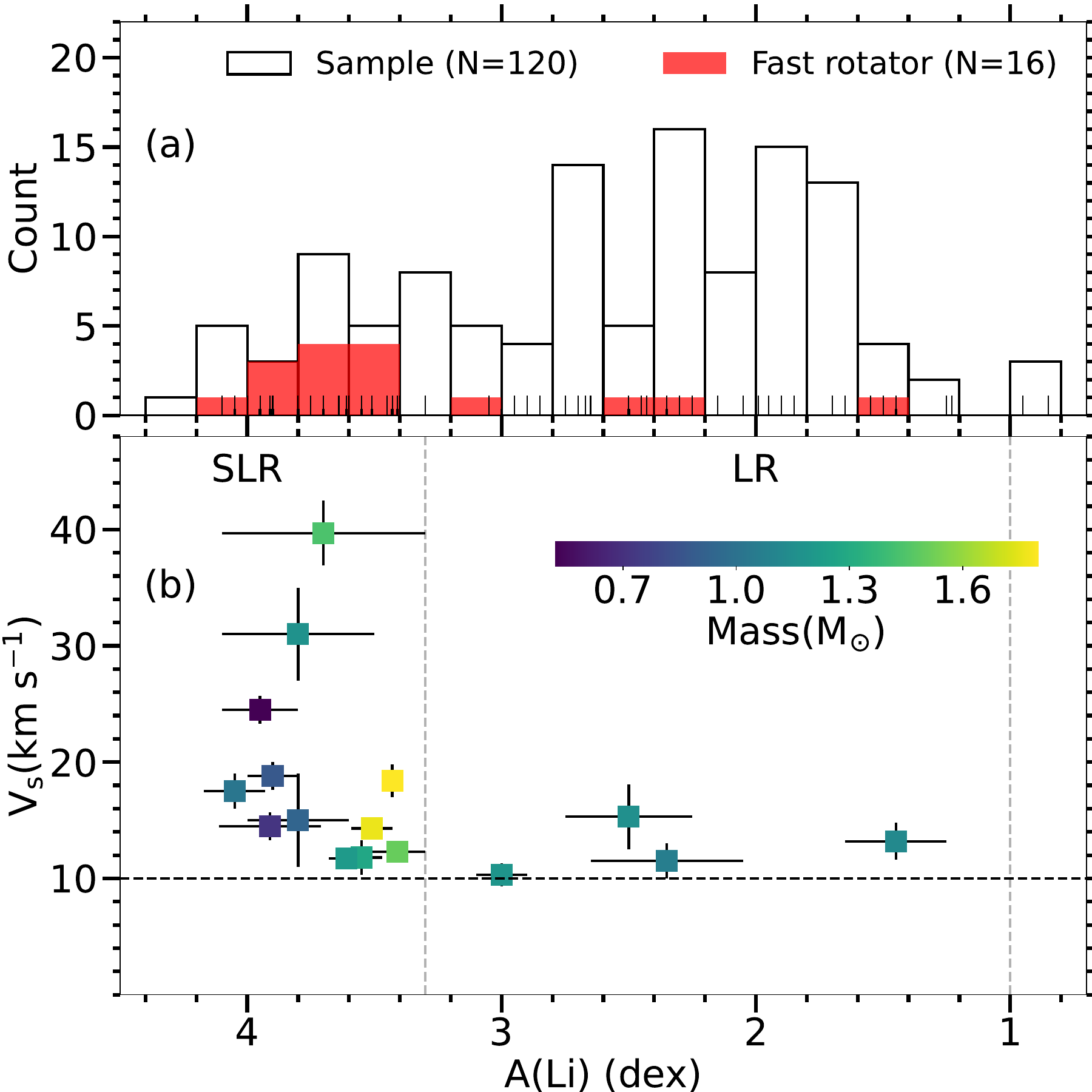}
    \caption{ Distribution of stars' surface rotation with lithium abundance for 16 red clump giants (top panel). The bottom panel is the plot of surface rotation against Li abundance. The colour bar indicates the mass of the stars. The dotted vertical line separates SLR giants (A(Li) $\geq$ 3.2~dex) from Li-rich giants (1.0$<$A(Li)$<$3.2). We find a moderate positive correlation (Pearson coefficient, r = 0.36) between A(Li) and V$_{\rm S}$. In SLR group, surface velocity is scattered with a standard deviation of 8.2~km s$^{-1}$ and in the LR group, stars are clustered with a standard deviation of 1.9~km s$^{-1}$.}
    \label{fig:alivsmass}
\end{figure}

\begin{figure*}[!ht]
    \centering
    \includegraphics[width=\textwidth]{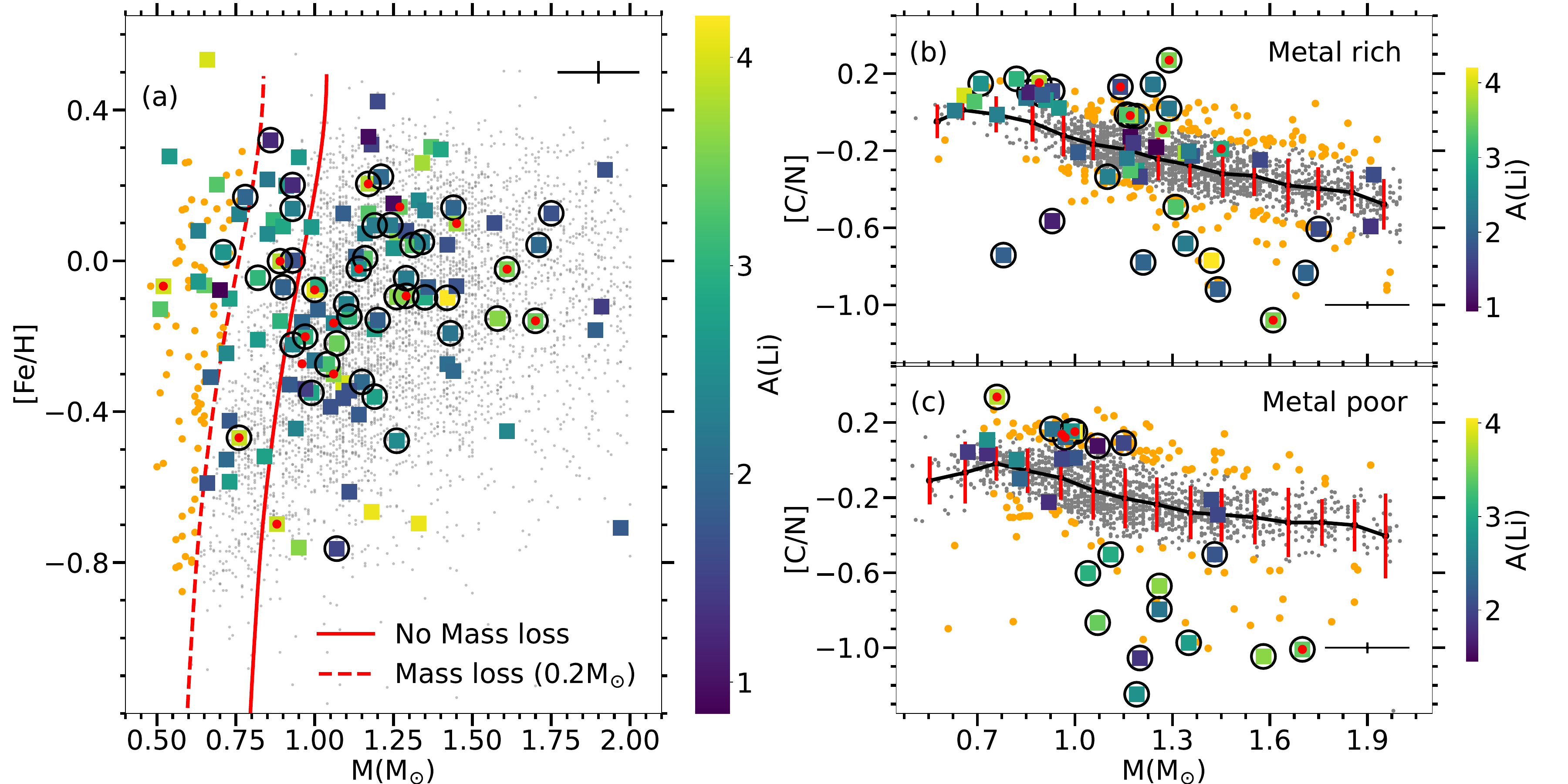}
    \caption{In the left panel, mass vs metallicity for the entire sample of 5227 stars is shown. Grey and orange symbols represent RC giants with no detectable Li line. The RC giants with A(Li) measured are shown as large coloured symbols. The dashed and solid red lines denote the minimum mass of RC giants without and with mass loss ($\sim$0.2~M$_{\odot}$), respectively. The orange symbols left of the dashed red line represent RCs with extremely low mass attributed to the past binary interactions \citep{Li2022, Matteuzzi2024}. Note, of the 89 extremely low-mass RC giants, 13 are Li-rich which is about 15~$\%$, a high occurrence of 7 times more than Li-rich giants among the RC sample. The encircled squares represent those with anomalous [C/N] (see right panel), and the red dotted squares are fast-spinning RC giants. On the right, plots of [C/N] and Mass are shown for two metallicities: (a) Metal-rich (Fe/H]$\ge$-0.1 dex) and (b) metal-poor [Fe/H]$<$-0.1 dex. The bulk of the RC giants follow the main trend, which is the average of [C/N] values in mass bins of 0.1~M$_{\odot}$. Red clump giants with [C/N] ratio 1.5~$\sigma$ away from the average trend on either side are outliers \citep{Bufanda2023}. A total of 77 Li-rich RC giants have [C/N]-ratio available. We found 43 Li-rich RC giants among the 351 outliers with anomalous [C/N], which is about 10 times more than finding Li-rich RC giants among the total sample.}
    \label{fig:massfehali}
\end{figure*}

The other possibility for high rotation among RCs is the external source, such as tidal synchronization in the binary system or the merger of a companion. We searched for evidence for the binary companion among the sample RC giants using the Gaia astrometry data (re-normalized unit weight error, RUWE) and variation in radial velocities {\it Vscatters} from APOGEE data. The presence of secondary can manifest in the form of an inflated value of Gaia re-normalized unit weight error (RUWE $\ge$ 1.4, \citep{Ziegler2020}). In our sample of 120 stars, none of the SLR giants have a RUWE value higher than 1.4, whereas only one LR giant has a RUWE value higher than 1.4. We did not find significant variation in radial velocities. The values of {\it Vscatters} from APOGEE indicate scatter in the radial velocity of stars observed by APOGEE multiple times. However, we note that most of the stars have radial velocity data for only 3 or 4 epochs. It would be difficult to conclude the binary status of the objects based on available data. The only exception is the KIC~11615224 with A(Li)= 3 dex, and has a Gaia value of RUWE = 2.65, suggesting a binary companion. The rest of the RC giants in the sample seem to be single. However, this doesn't rule out their past association with the binary interaction/merger. The transfer of angular momentum during the companion merger may result in high rotation. There are about ~41$\%$ of the low mass G- and K-type main sequence stars that are found to be in binary companions \citep{Raghavan2010}. The possibility of interaction or eventual merger of companions increases as stars evolve and increase in size \citep{Li2022}. Post-merger stars appear as single stars with altered physical and chemical properties like increased surface rotation, altered surface composition and extremely low mass due to excess mass-loss during or immediately after the merger event  \citep{Bufanda2023, Matteuzzi2024, Rui2024}. 

It is quite interesting to see that most (75~$\%$) of the high rotation RC giants are super Li-rich, and all of them are Li-rich. The plot of V$_{\rm rot}$ versus A(Li) (Figure~\ref{fig:alivsmass}(a)) shows a rapid decrease in the distribution of surface rotation with decreasing A(Li) value. It implies that both the Li and rotation enhancements are probably associated with an external event. It was earlier demonstrated that the high A(Li) or super Li-rich giants are younger RCs, i.e. enhancement of Li occurred relatively recently compared to RCs with lower Li values \citep{singh2021}. As shown in Fig~\ref{fig:alivsmass}(b), the extremely high Li-rich (mean A(Li)$\sim$3.9) giants have much higher V$_{\rm rot}$ ranging from 14 to 40~km s$^{-1}$. Giants with mean A(Li)$\sim$3.4~dex show V$_{\rm rot}$ in the range of 11--19~km s$^{-1}$ and giants with A(Li) $<$ 3.2~dex have V$_{\rm rot}$ of about 11~km s$^{-1}$. This shows that Li-richness and high surface rotation among RCs are rapidly decreasing from their initial values, and they seem transient. The correlation is an indication that Li enhancement is probably associated with external events such as mergers. 

The merger scenario also finds indirect evidence of the presence of a high proportion of Li-rich giants among extremely low-mass RCs. In Fig~\ref{fig:massfehali}(a), entire sample RC giants are shown in a plot of metallicity versus stellar mass to identify extremely low-mass RCs. Some of the masses (M $\leq$ 0.7~M$_{\odot}$) are such that they wouldn't have evolved to the RCs phase, and their ages exceed the age of the universe if we account for the expected mass loss (~0.2~M$_{\odot}$) during the evolution of RGB \citep{Miglio2021}.
% The masses ($\leq$ 0.7~M$_{\odot}$) of the giants are such that they would be much older than the age of the universe even if we account for the expected mass loss (~0.2~M$_{\odot}$) during the evolution of RGB \citep{Miglio2021}.
We found 13 Li-rich giants among the extremely low mass regime of the plot (Fig~\ref{fig:massfehali}(a)), which is about a factor of five larger than the Li-rich giants' share among the RCs. This means the extremely low-mass RCs might have experienced excess mass loss apart from the normal mass loss (0.2~M$_{\odot}$) during their evolution \citep{Miglio2021}. Recent studies show that the extremely low mass RCs might be the result of excess mass loss during past mergers or binary interaction \citep{Li2022, Matteuzzi2024}. Finding a one-to-one correlation between high Li, extremely low mass and high rotation is difficult as their enhancement may vary from star to star, subject to the complexity of interaction and merger scenario coupled with the mixing and the level of lithium abundance during the He-flash. The final mass of RC giants is subject to the level of mass loss during the merger events, which depends on the mass of the companion, proximity and level of mass transfer, etc. Studies also suggest that stars that have undergone binary interaction or merger are expected to have anomalous [C/N] ratios for their mass \citep{Bufanda2023}. Though the exact physical mechanism is not well understood, these RC giants are found to be outliers from the bulk of the sample giants in a plot of [C/N] versus mass (Fig~\ref{fig:massfehali}(b \& c)). These lie at about 1.5~$\sigma$ from the mean of the bulk of the RC giants. Only 12 out of 16 rapidly rotating Li-rich RC giants have [C/N] values; interestingly, 10 out of 12 high surface rotation RC giants belong to this [C/N] anomalous group. We found that finding Li-rich RC giants among the outliers is about a factor of ten more likely compared to finding them among the RC giants. 

\section{Conclusions}
We analyzed a large sample of red clump giants using spectroscopic and photometric data. We found a correlation between Li and surface rotation among the RC giants for which we could measure actual surface rotation using stellar spots. Of the 30 super Li-rich giants in our sample, we found that 40~$\%$ of them are rapid rotators. Among the 16 rapid rotators, 12 are super Li-rich and four are Li-rich. The association of very high surface rotation with most of the super Li-rich giants and their spin-down with decreasing Li implies that the two properties are acquired very recently, either due to the merger-induced He-flash or due to merger events just before or after the He-flash. While the He-flash may be responsible for the rapid mixing of stars' Li-rich material with the outer surface layers, the merger events might have contributed to the large observed surface rotation, probably aiding the mixing process. Another piece of evidence is the high proportion of Li-rich giants among the extremely low-mass red clump stars. The very low mass is attributed to excess mass loss during the interaction or merger events \citep{Li2022, Matteuzzi2024}. We found a factor of seven more likely for finding Li-rich giants among the extremely low mass RCs compared to among the low mass RCs. We also note another observation of the sample data, which is the presence of RC giants with extremely low or high values of [C/N/]. Some studies attribute this to the past histories of binary interaction or merger events. We found that finding Li-rich RC giants among the sample of RC giants with anomalous [C/N] values is more likely, by a factor of 10, compared to among the RC sample. These results pose questions about whether the merger or binary interactions are necessary for the high Li abundance seen in the RC giants. 

\section{Acknowledgment}
This study is supported by the National Natural Science Foundation of China under grant No. 11988101, and the National Key R\&D Program of China No. 2024YFA1611900. We thank Prof. Hong-Liang Yan, Haining Li and JC Pandey for helpful discussion on the manuscript. Guoshoujing Telescope (the Large Sky Area Multi-Object Fibre Spectroscopic Telescope, LAMOST) is a National Major Scientific Project built by the Chinese Academy of Sciences. LAMOST is operated and managed by the National Astronomical Observatories, Chinese Academy of Sciences. This work presents results from the European Space Agency (ESA) space mission Gaia. Gaia data are being processed by the Gaia Data Processing and Analysis Consortium (DPAC). Funding for the DPAC is provided by national institutions, in particular the institutions participating in the Gaia MultiLateral Agreement (MLA). The Gaia mission website is https://www.cosmos.esa.int/gaia. This paper includes data observed by the NASA Kepler space mission and retrieved from the MAST data archive at the Space Telescope Science Institute. Funding for the Kepler mission is provided by the NASA Science Mission Directorate.  We thank the anonymous referee for useful suggestions for greater clarity.

% \bibliography{ref}  

\bibliographystyle{aasjournal}
\end{document}